# Air Quality and Greenhouse Gas Emissions Assessment of Data Centers in Texas: Quantifying Impacts and Environmental Tradeoffs

Ebrahim Eslami[*,**]
[*] EnviroPilot.ai, Houston, TX
[**] (formerly at) Houston Advanced Research Center (HARC), The Woodlands, TX

## Summary:

This study comprehensively assessed air quality (AQ) and greenhouse gas (GHG) emissions associated with the rapid expansion of data centers in Texas. Recognizing Texas as a major data center hub due to its infrastructure, electricity market, and favorable business conditions, the study distinctly separated AQ impacts from GHG emissions to clarify their different sources, regulatory frameworks, and mitigation strategies.

The analysis highlighted substantial GHG emissions, primarily from electricity consumption and cooling systems, as the dominant contributor. Operational electricity use in a standard 10-megawatt data center was estimated to generate approximately 37,668 metric tons of $CO_2$ annually. Additionally, embodied emissions from construction materials and IT equipment significantly contribute to the total lifecycle carbon footprint.

Local AQ impacts, often overlooked in existing literature, were closely examined. Diesel-powered backup generators, construction equipment, and employee commuting were identified as notable sources of criteria pollutants such as nitrogen oxides ($NO_x$) and particulate matter (PM), particularly in urban regions already facing air quality challenges. For instance, generator testing alone can emit around 12 metric tons of $NO_x$ annually per large data center, exacerbating ozone pollution in areas such as Houston and Dallas-Fort Worth.

Several mitigation strategies were discussed, including advanced cooling technologies, renewable energy integration, cleaner backup power solutions such as fuel cells and battery storage, sustainable construction practices, and comprehensive emission reporting frameworks. Case studies from Texas (e.g., CyrusOne, Microsoft, Digital Realty) and international best practices provided practical examples of successful emission reduction approaches.

Predictive modeling based on ERCOT's 2025 Long-Term Load Forecast shows that electricity demand from data centers in Texas is projected to grow significantly by 2030. Utility-submitted forecasts estimate up to around 78 gigawatts of new data center load by the end of the decade, with ERCOT adjusting that figure to approximately 39 gigawatts to reflect historical implementation rates. If this expansion occurs without targeted interventions, associated greenhouse gas emissions from operational electricity use alone could range between 170 to 205 million metric tons of $CO_2$ per year, depending on realized capacity, facility efficiency, and grid carbon intensity. However, with aggressive adoption of renewable energy procurement, advanced cooling systems, and cleaner backup power technologies, emissions could be reduced by 50 to 80%, potentially avoiding 85 to 165 million metric tons of annual $CO_2$ emissions across Texas by 2030. These findings highlight both the scale of the environmental challenge and the critical role that proactive technological and policy measures can play in shaping sustainable digital infrastructure growth.

The paper identified critical research and policy gaps, emphasizing the need for cumulative air dispersion modeling, standardized emissions reporting, and AQ-specific regulatory frameworks. It concluded with actionable recommendations for policymakers and industry stakeholders, advocating mandatory efficiency standards, renewable energy mandates, AQ-focused regulations, and enhanced transparency through emissions disclosures. Ultimately, proactive adoption of these strategies can balance Texas's digital infrastructure growth with essential environmental and community health protections, ensuring sustainability in the long term.

**Keywords:** Data centers; Texas; Air quality; Greenhouse gas emissions; Lifecycle assessment; Diesel generators; Embodied carbon; Cooling systems; Renewable energy; Scope 1 emissions; Scope 2 emissions; Scope 3 emissions; Energy infrastructure; Environmental impact; ICT sector; Sustainability.

Table of Contents

# 1. Introduction

The rapid increase in numbers of digital technologies has driven exponential growth in data center construction and electricity use. Globally, data centers consumed approximately 220 to 320 terawatt-hours (TWh) of electricity in 2021, accounting for 0.9 to 1.3 percent of global final electricity demand (IEA, 2022). In the United States, data centers consumed about 2 percent of the national electricity supply, with the majority concentrated in high-demand regions such as Northern Virginia, Silicon Valley, and Texas (Wang et al., 2023). Within Texas, the Dallas–Fort Worth metroplex alone hosts over 400 megawatts (MW) of commissioned colocation capacity, placing it among the top three data center hubs in North America (CBRE, 2023).

The environmental impact of this infrastructure extends across the full life cycle of data center construction, operation, and decommissioning. Greenhouse gas (GHG) emissions are a major concern, particularly Scope 2 emissions from electricity consumption. For a typical facility operating at 10 MW around the clock, annual electricity use exceeds 87,000 megawatt-hours (MWh), which, if powered by a natural gas-dominant grid like ERCOT's, corresponds to over 30,000 metric tons of carbon dioxide equivalent ($CO_{2e}$) per year (Hasan et al., 2022). In areas where coal-fired generation is still active, such as parts of East Texas, the emissions intensity can exceed 0.55 metric tons $CO_{2e}$ per MWh (EPA eGRID, 2022).

In addition to electricity-based emissions, data centers also generate Scope 1 emissions through on-site combustion sources, primarily diesel backup generators. A single 2.5 MW diesel generator emits over 5 metric tons of nitrogen oxides ($NO_x$) and 0.3 metric tons of particulate matter ($PM_{10}$) during 100 hours of annual testing, based on EPA AP-42 emission factors. Many hyperscale facilities operate multiple generators, which can collectively emit several dozen tons of criteria pollutants each year, especially when tested in parallel. In Texas, where generator testing is often performed during hot summer months to align with peak demand reliability protocols, these emissions can coincide with meteorological conditions that exacerbate ozone formation and pollutant dispersion constraints (Mughal et al., 2023; TCEQ, 2022).

Scope 3 emissions, which include embedded carbon from construction materials, server manufacturing, refrigerants, and logistics, often rival or exceed operational emissions. For example, life cycle assessments indicate that server manufacturing alone can account for 20 to 30 percent of a data center's 15-year GHG footprint, depending on hardware refresh rates (Alva et al., 2022; Sharma et al., 2023). Recent work by Stobbe et al. (2023) emphasized that shifting IT workloads to cloud providers may obscure these upstream emissions under traditional accounting frameworks, since customer emissions become categorized as Scope 3 under cloud vendor inventories.

While the carbon impacts of data centers have been increasingly quantified in recent years, the associated air quality (AQ) impacts remain insufficiently characterized, particularly at the local and regional level. Diesel generators emit not only $CO_2$ but also $NO_x$, volatile organic compounds (VOCs), carbon monoxide (CO), and particulate matter, all of which are regulated under the Clean Air Act. The Dallas–Fort Worth region is currently designated as nonattainment for ozone under the 2015 National Ambient Air Quality Standards (NAAQS). In such regions, even minor increases in $NO_x$ can contribute to violations of federal standards due to the region's

NO$_x$-limited photochemical regime. Generator emissions, though episodic, are not negligible. A cluster of ten data centers, each with five backup generators operating for 100 hours per year, can release over 250 metric tons of NO$_x$ annually, enough to trigger regulatory scrutiny or permit modifications if aggregated as a stationary source complex (Qureshi et al., 2022; Nayyar et al., 2021).

Cooling systems further complicate both GHG and AQ assessments. Direct refrigerant leakage can release hydrofluorocarbons (HFCs), which have global warming potentials (GWPs) between 1,300 and 3,900 times that of CO$_2$. For instance, R-410A, a common refrigerant in data center HVAC systems, has a GWP of 2,088. If 5 percent of a 1,000-pound charge leaks annually across a facility, the resulting emissions are equivalent to over 100 metric tons of CO$_2$e. Indirectly, the electricity required for cooling increases both GHG emissions and power plant-related air pollutant emissions, particularly during peak summer periods when air conditioning loads dominate grid profiles (Kamali & Hewage, 2023).

Despite these impacts, air quality regulation of data centers in Texas remains fragmented. Backup generators are typically permitted as emergency units and may be exempt from detailed dispersion modeling under the assumption of infrequent use. However, in practice, cumulative testing emissions, combined with load shedding or grid support contracts, can lead to routine non-emergency operation. TCEQ permitting guidance generally does not require ambient air quality analysis for backup units unless trigger thresholds are exceeded, which may fail to capture cumulative impacts in densely clustered industrial parks such as those in Houston's Energy Corridor or Richardson's Telecom Corridor (TCEQ, 2023; Mughal et al., 2023).

The scientific literature has advanced a number of mitigation strategies, particularly in the area of GHG emissions. These include carbon-aware workload scheduling, dynamic load shifting between geographic regions with cleaner electricity, power purchase agreements (PPAs) with renewable providers, and server energy optimization algorithms (Zhao et al., 2023). Less attention has been paid to technologies or strategies aimed at minimizing air pollutant emissions. Potential approaches include substituting diesel generators with natural gas or hydrogen-based systems, deploying battery energy storage, or scheduling generator testing during periods of favorable atmospheric dispersion.

This paper aims to provide a comprehensive and quantitative assessment of both greenhouse gas and air quality emissions associated with data center development and operation in Texas. It systematically distinguishes between direct and indirect emission sources, assesses their spatial and temporal variability, and evaluates regulatory frameworks and mitigation options. A particular emphasis is placed on the underexplored air quality dimension, given its implications for local health outcomes, permitting, and environmental justice. By treating GHG and AQ domains in parallel but distinct analytical tracks, this study offers a more complete scientific basis for environmental planning in the rapidly expanding digital economy of Texas.

**1.1 Overview of Data Centers in Texas:**

Texas has emerged as one of the fastest-growing regions for data center development in the United States, driven by a combination of factors including low electricity prices, abundant land, favorable regulatory conditions, business-friendly tax incentives, and access to robust fiber infrastructure. According to ERCOT's 2025 Long-Term Load Forecast, utility-submitted plans suggest that up to 77,965 megawatts of new data center electricity demand could materialize by 2030, with a realistically adjusted projection of approximately 38,878 megawatts based on historical fulfillment trends. The Dallas-Fort Worth metroplex remains the state's dominant data center hub, accounting for a significant share of current and planned capacity, followed by active clusters in Austin, San Antonio, and Houston. As of 2023, Dallas-Fort Worth alone hosted over 590 megawatts of commissioned multi-tenant data center capacity, placing it among the top three data center markets in North America in terms of both floor space and electrical load (CBRE, 2023). This rapid growth trajectory positions Texas as a key contributor to the national data infrastructure buildout while also making it a critical focal point for addressing the environmental impacts of large-scale digital operations.

Electricity supply is a key driver in the siting and operation of data centers. Texas operates its own independent power grid through the Electric Reliability Council of Texas (ERCOT). The ERCOT grid serves approximately 90 percent of the state's electric load. Its generation mix includes natural gas (42 percent), wind (24 percent), coal (17 percent), solar (6 percent), and nuclear (10 percent), based on 2022 data. Because of this mix, the carbon intensity of grid electricity in Texas averages about 0.43 metric tons of $CO_2$ per megawatt-hour (EPA eGRID, 2022). However, this value can vary significantly depending on location and time of day. For example, during midday in West Texas, solar and wind can dominate generation, while in East Texas, coal and gas remain major contributors.

The size and configuration of data centers in Texas vary widely. Enterprise and colocation facilities typically range from 5 MW to over 50 MW of connected capacity. Hyperscale facilities built by companies like Meta, Microsoft, and Google often exceed 100 MW, with campuses spread across hundreds of acres. Smaller edge data centers, serving local caching or IoT functions, operate in the 0.5 to 2 MW range. Despite their differences in scale, all types rely on continuous electricity supply, redundant backup systems, and climate control to maintain high availability.

Texas offers a combination of financial and regulatory incentives that have contributed to its rapid growth as a data center hub. At the state level, the Texas Data Center Exemption provides sales and use tax exemptions for facilities investing at least $200 million and creating a minimum of 20 jobs, covering items such as servers, cooling systems, generators, and electricity use (Texas Comptroller of Public Accounts, 2023). Additional property tax abatements are available through local agreements such as Chapter 312, often reducing taxes by up to 80 percent (Griffith & Williams, 2021). Moreover, the Texas Commission on Environmental Quality (TCEQ) allows emergency diesel generators to operate under a permit-by-rule (30 TAC §106.511), reducing permitting burdens for backup power infrastructure (TCEQ, n.d.).

Data centers in Texas are typically designed with N+1 or 2N redundancy for both power and cooling systems, meaning that at least one backup unit is available for every critical system (Uptime Institute, 2023). Most facilities deploy multiple diesel generators, each typically ranging from 1 to 3 MW, to meet full site load during utility outages (Hasan et al., 2022). In a 30 MW data center, this often results in 10 to 15 generators, each equipped with large aboveground diesel storage tanks and automatic transfer switches (Kamali & Hewage, 2023). These units are not idle; monthly testing and routine maintenance are standard industry practice, resulting in 50 to 150 operating hours per generator annually, even in the absence of grid failures (EPA, 2021).

Permitting backup generator systems falls under the authority of the TCEQ. These units are typically classified as emergency sources and are often authorized under a permit-by-rule, which exempts them from detailed air dispersion modeling as long as they operate below defined usage thresholds (TCEQ, n.d.). While this approach simplifies compliance for individual facilities, it does not account for the cumulative impacts of multiple data centers operating within proximity such as in industrial parks or technology corridors; areas where aggregated emissions of $NO_x$ and PM may contribute to local air quality degradation, especially in nonattainment regions.

Cooling infrastructure is a major contributor to energy consumption in data center operations. Most large facilities in Texas rely on air-cooled chillers or direct expansion systems, although liquid cooling is becoming more common in high-density computing environments (Kamali & Hewage, 2023). Cooling systems can account for 20 to 40 percent of a data center's total energy use, with the exact share depending on factors such as ambient temperature, humidity, and server utilization (Hasan et al., 2022). The climate in Texas, particularly during extended summer periods, increases cooling demand relative to regions with more temperate weather, such as the Pacific Northwest. Consequently, Power Usage Effectiveness (PUE) values for data centers in Texas typically range from 1.3 to 1.5 under optimized design and operational conditions (ASHRAE, 2021).

The siting of data centers relative to urban populations, vulnerable communities, and ozone nonattainment regions is an important consideration for both public health and regulatory oversight. In the Houston metropolitan area, many data centers are located along Beltway 8 and in the Energy Corridor, where they are adjacent to communities already burdened by industrial air pollution (TCEQ, 2023). Similarly, in the Dallas-Fort Worth region, clusters of large-scale facilities have developed near residential areas in cities such as Plano, Richardson, and Irving (EPA, 2022). In these locations, emissions from diesel backup generators can exacerbate existing air quality challenges, particularly during the ozone season or periods of atmospheric stagnation when pollutant dispersion is limited. These cumulative exposures may be of concern in areas already designated as nonattainment for ground-level ozone under the NAAQS.

This section shows that Texas provides a favorable environment for data center development due to its infrastructure, energy market, and policies. However, the same characteristics that make Texas attractive for this industry, such as lenient permitting and fossil fuel reliance, also increase the potential for environmental impacts. The next sections will provide a deeper analysis of those impacts, starting with a full accounting of greenhouse gas emissions across the data center life cycle, followed by air quality emission sources and their implications for public health and compliance.

# 2. Literature Review

## 2.1 Current Trends in Data Center Development in Texas

Data center development in Texas has accelerated over the past decade, driven by increasing demand for cloud computing, artificial intelligence, and high-speed content delivery. According to commercial market reports, the Dallas–Fort Worth metroplex is among the top three largest data center hubs in North America, with more than 400 MW of commissioned colocation capacity as of 2023. Houston, Austin, and San Antonio are also experiencing rapid expansion, fueled by enterprise demand and the availability of competitive electricity pricing (Wang et al., 2023).

The state's independent electricity grid, operated by the ERCOT, is a major draw for operators seeking control over power sourcing. In 2022, the ERCOT grid's generation mix included 42 percent natural gas, 24 percent wind, 17 percent coal, and 6 percent solar, with the remaining share from nuclear (EPA eGRID, 2022). Because of this mix, the carbon intensity of electricity in Texas remains relatively high compared to regions with a greater share of renewables.

Data center siting patterns in Texas show clustering in urban and suburban areas with existing fiber infrastructure and low-cost land. For example, the city of Richardson, north of Dallas, houses dozens of hyperscale and colocation facilities in close proximity, often within 500 meters of residential zones. Similarly, Houston's Energy Corridor and areas along Beltway 8 are home to multiple enterprise and disaster-recovery centers. These patterns raise concerns about cumulative local emissions from diesel backup generators and cooling systems, particularly in ozone nonattainment areas such as Dallas–Fort Worth and Houston.

## 2.2 Previous Studies on Data Centers' Impact on AQ and GHG Emissions

A growing body of research has assessed the climate impacts of data centers, although fewer studies have examined their local air quality consequences. Most GHG-focused studies emphasize Scope 2 emissions, which dominate due to the heavy electricity demands of servers, cooling systems, and infrastructure. Hasan et al. (2022) and Sharma et al. (2023) report that operational emissions typically account for more than 60 percent of a facility's total lifecycle carbon footprint. These emissions vary based on data center location, the local grid's fuel mix, and the energy efficiency of installed equipment.

Lifecycle assessments (LCA) of data centers reveal that Scope 3 emissions can also be substantial. Alva et al. (2022) estimate that the embedded carbon from server manufacturing, construction materials, and supply chains contributes 25 to 35 percent of total lifecycle GHGs, depending on hardware replacement intervals. Data centers with shorter refresh cycles or extensive use of high-performance computing clusters may exhibit even higher Scope 3 ratios.

Kamali and Hewage (2023) highlight that cooling systems can contribute both directly and indirectly to emissions. Direct emissions occur through the leakage of refrigerants with high global warming potential (e.g., R-134a, R-410A), while indirect emissions result from increased electricity use, especially during summer months in warm climates like Texas. Cooling can account for up to 40 percent of a facility's energy use, particularly in air-cooled systems without economizer cycles.

Compared to GHG studies, relatively few have addressed air quality impacts in detail. However, several papers document the emissions of criteria pollutants from backup generators. Qureshi et al. (2022) and Nayyar et al. (2021) model emissions from diesel generator testing at large-scale data centers and report annual emissions ranging from 10 to 50 tons of $NO_x$ per site, depending on testing hours and generator configuration. These emissions are particularly concerning in ozone-sensitive regions like North Texas, where $NO_x$ contributes to local exceedance of the NAAQS.

Other studies, such as those by Stobbe et al. (2023), caution that cloud-based computing may mask environmental burdens by shifting them from users to providers. Their work suggests that more transparent carbon accounting methods are needed to avoid underreporting of cloud-based emissions, especially as AI and big data applications grow.

A number of modeling studies also propose mitigation strategies, such as carbon-aware workload shifting, geographical load balancing based on real-time grid emissions, and use of on-site renewables. Zhao et al. (2023) demonstrate that shifting workloads from carbon-intensive regions to cleaner grids can reduce Scope 2 emissions by up to 20 percent without hardware upgrades. However, such strategies often require advanced coordination and may have unintended consequences on local power loads and grid stability.

Despite growing attention to the environmental footprint of data centers, important research gaps remain. First, air quality impacts are often overlooked or treated only qualitatively in the literature. Most existing studies either focus exclusively on GHGs or combine AQ and GHG emissions without distinguishing their regulatory, spatial, or temporal implications. This is a critical limitation, particularly in Texas where ozone nonattainment status, local permitting policies, and cumulative community exposures demand a more detailed understanding of criteria pollutant emissions. Second, there is limited regional analysis of how backup generator emissions interact with local meteorology, atmospheric chemistry, and neighborhood vulnerability. Few studies employ dispersion modeling or health risk assessment to quantify potential exposures near clustered data center campuses. Third, while lifecycle GHG emissions have been studied extensively at global or national scales, very few assessments disaggregate impacts by component (e.g., servers vs. construction materials) or by lifecycle stage (e.g., manufacturing vs. disposal) in a Texas-specific context. Finally, current mitigation studies often prioritize technological innovation (e.g., load shifting, renewable sourcing) over operational changes that may yield near-term reductions in both GHG and AQ impacts, such as rescheduling generator tests or using low-emission fuels.

## 3. Data Center Construction and Its Impact

### 3.1 Materials Used in Construction

Construction materials significantly affect the environmental footprint of data centers. Key materials include reinforced concrete, structural steel, aluminum, glass, insulation, and copper wiring. Concrete and steel represent the largest portion, accounting for about 70-85% of total construction-related emissions (Alva et al., 2022).

Concrete production involves calcination of limestone and fossil fuel combustion, resulting in high $CO_2$ emissions. Typically, manufacturing one ton of cement releases approximately 0.9 tons of $CO_2$. A standard 10-megawatt (MW) data center might utilize between 5,000 and 10,000 cubic meters of concrete, leading to an estimated emission of around 1,000-2,000 metric tons of $CO_2$. Structural steel, another major component, emits approximately 1.9 tons of $CO_2$ per ton produced. A data center of similar size typically requires around 500-1,000 tons of structural steel, thus contributing an additional 950-1,900 metric tons of $CO_2$ (Hasan et al., 2022; Sharma et al., 2023).

Manufacturing these materials also generates air pollutants such as PM, $NO_x$, sulfur dioxide ($SO_2$), and VOCs. Although these emissions occur off-site, they affect air quality in nearby communities, especially those located close to cement plants and steel mills.

**3.2 Energy Consumption During Construction**

Construction activities for data centers involve extensive use of diesel-powered equipment, including cranes, excavators, loaders, bulldozers, concrete mixers, and generators. These machines consume significant amounts of diesel fuel, resulting in direct emissions of greenhouse gases and air pollutants.

The U.S. EPA estimates that a medium-sized construction project consumes approximately 200,000-500,000 liters of diesel fuel, producing 500-1,300 metric tons of $CO_2$ (EPA NONROAD, 2021). Construction activities also emit substantial amounts of $NO_x$, PM, and CO from heavy machinery, as summarized below:

- $NO_x$: Approximately 3-5 metric tons per 100,000 liters of diesel fuel consumed.
- PM: Approximately 0.2-0.5 metric tons per 100,000 liters.
- CO: Approximately 0.6-1.0 metric tons per 100,000 liters (EPA NONROAD, 2021).

Considering typical fuel consumption during data center construction, total emissions can be significant, especially in densely populated urban or suburban locations. Table 1 presents estimated emissions from construction activities associated with a typical 10 MW data center, including both embodied and direct sources.

Table 1. Estimated Emissions from Construction Activities for a 10 MW Data Center

| Source/Activity | GHG Emissions (metric tons $CO_2$) | $NO_x$ (metric tons) | PM (metric tons) | CO (metric tons) |
|---|---|---|---|---|
| Concrete Production | 1,000-2,000 | - | - | - |
| Steel Production | 950-1,900 | - | - | - |
| Diesel Fuel Combustion (on-site machinery) | 500-1,300 | 6-25 | 0.4-2.5 | 1-5 |
| Material Transportation (diesel trucks) | 100-300 | 2–5 | 0.1-0.5 | 0.5-1.5 |
| Total | 2,550-5,500 | 8-30 | 0.5–3.0 | 1.5-6.5 |

(Sources: Hasan et al., 2022; Sharma et al., 2023; EPA NONROAD, 2021)

### 3.3 Urban Planning and Location Considerations

The site chosen for data centers significantly influences environmental impacts. Siting facilities in undeveloped rural areas requires new infrastructure development, including roads, electrical transmission lines, water and sewer systems, which increases overall emissions. Alternatively, locating data centers in urban industrial zones may decrease the need for new infrastructure but could exacerbate local air quality issues due to cumulative emissions from multiple sources.

In Texas, many data centers are clustered in suburban areas or near urban industrial zones. The Dallas–Fort Worth metroplex, particularly Richardson and Plano, hosts dense concentrations of facilities. Similarly, Houston's Energy Corridor along Beltway 8 includes numerous data centers close to residential neighborhoods. These siting patterns raise concerns regarding cumulative air quality impacts and the adequacy of existing regulatory oversight.

Currently, Texas lacks detailed permitting requirements for data center construction regarding local air quality impacts. Unlike power plants or industrial facilities, data center construction rarely triggers Environmental Impact Statements (EIS). Furthermore, air dispersion modeling and cumulative impact analyses are not mandatory for typical data center construction permits, creating a potential regulatory gap.

### 3.4 Research and Policy Gaps

A significant research gap exists regarding detailed quantification of emissions at each construction stage, particularly concerning air quality pollutants. Most available studies primarily focus on greenhouse gases and rarely differentiate emissions based on specific construction activities or machinery types. Furthermore, construction-phase air quality modeling, particularly dispersion modeling of $NO_x$ and PM emissions, is virtually absent from the literature.

On the policy side, Texas currently lacks stringent regulations for mitigating air pollutant emissions during construction. Adoption of cleaner, more efficient equipment (such as Tier 4 diesel engines with advanced emission controls) is not mandatory statewide for private-sector data center construction. Without explicit regulatory incentives or mandates, widespread adoption of cleaner construction practices remains uncertain.

## 4. Operational Emissions

The operational phase is the longest period in a data center's lifecycle and contributes substantially to both GHG emissions and AQ impacts. These emissions are primarily driven by electricity use, cooling systems, and backup power operations, each of which has distinct environmental implications.

### 4.1 Energy Consumption

**Direct Emissions from Electricity Use**

Data centers rely predominantly on electricity drawn from local power grids. Texas operates its own independent grid, known as the ERCOT, which covers approximately 90% of the state. ERCOT's power generation in 2022 comprised primarily natural gas (42%), wind power (24%), coal (17%), nuclear (10%), and solar (6%) (EPA eGRID, 2022). This energy mix yields an average emissions intensity of approximately 0.43 metric tons of $CO_2$ per MWh.

For context, a typical 10 megawatt (MW) data center running continuously consumes roughly 87,600 MWh annually (calculated as 10 MW multiplied by 24 hours/day and 365 days/year). Given ERCOT's average emission factor, this equates to approximately 37,668 metric tons of $CO_2$ emissions per year from electricity alone, a substantial contribution when considering the numerous data centers across Texas.

The type of power source directly influences these emissions. Facilities drawing electricity predominantly from renewable sources, such as wind or solar, can significantly lower operational carbon emissions. Conversely, facilities served mainly by coal or natural gas have notably higher emissions intensities. Given the geographical variation in Texas's power generation, the location of a data center within the state critically influences its environmental footprint.

**Indirect Emissions from the Energy Supply Chain**

Beyond direct electricity use, additional emissions arise indirectly from fuel extraction, processing, and transportation within the electricity supply chain. Upstream activities add approximately 5 to 10 percent to the direct emissions of electricity production, and transmission losses within the grid further increase this figure by an estimated 6 to 7 percent. While often excluded from facility-level emissions inventories, these indirect emissions contribute meaningfully to the data center's full lifecycle impact and should be accounted for when evaluating comprehensive environmental strategies (Hasan et al., 2022).

## 4.2 Cooling Systems

Cooling systems are critical components of data centers, especially in Texas, where high summer temperatures elevate cooling demand significantly. These systems generally consume between 20 and 40 percent of the facility's total electricity, making them a key source of indirect emissions. Three primary cooling technologies are used: air-cooled chillers, water-cooled chillers, and direct expansion (DX) systems.

Air-cooled chillers are the most common in Texas, offering a balance of simplicity and efficiency, with typical Power Usage Effectiveness (PUE) values ranging from 1.3 to 1.5. Water-cooled chillers are more energy-efficient (PUE typically between 1.2 and 1.3), but their widespread use is restricted by water availability, especially in drought-prone areas. DX systems, often found in smaller data centers, are less efficient, frequently having PUE values exceeding 1.5, resulting in higher electricity consumption and associated emissions (Kamali & Hewage, 2023).

Cooling systems also utilize refrigerants, which pose direct greenhouse gas emission risks due to leakage. Refrigerants such as R-410A (GWP of 2,088) and R-134a (GWP of 1,430) are common in data center cooling applications. Annual leakage rates typically range from 2 to 5 percent of the total refrigerant inventory. For example, a large data center cooling system containing approximately 1,000 kg of R-410A could annually leak between 20 and 50 kg, equating to about 42 to 104 metric tons of $CO_2$-equivalent emissions. Given these impacts, refrigerant selection and leak management strategies present crucial opportunities to reduce GHG emissions at the operational level. Table 2 compares the energy use, emissions, and refrigerant leakage across common cooling system types for a 10 MW data center under typical Texas climate conditions.

Table 2. Cooling System Comparison for a 10 MW Data Center

| Cooling System | Typical PUE | Annual Electricity (MWh) | Annual $CO_2$ from Electricity (metric tons) | Annual Refrigerant Leakage (metric tons $CO_2$e) |
|---|---|---|---|---|
| Air-cooled | 1.3–1.5 | 26,280–43,800 | 11,300–18,800 | 42–104 |
| Water-cooled | 1.2–1.3 | 17,520–26,280 | 7,500–11,300 | 42–104 |
| DX systems | >1.5 | >43,800 | >18,800 | 42–104 |

(Sources: EPA eGRID, 2022; Kamali & Hewage, 2023)

## 4.3 Backup Power Systems

Data centers require highly reliable power supplies, typically employing diesel-powered backup generators to maintain continuous operations during grid outages. These generators are usually tested monthly, resulting in 50 to 150 hours of annual operation per generator, significantly contributing to local AQ issues.

A single 2.5 MW diesel generator tested for 100 hours annually consumes about 50,000 liters of diesel, producing approximately 130 metric tons of $CO_2$. Additionally, diesel combustion generates significant quantities of local air pollutants such as $NO_x$, PM, and CO. Using typical emission factors from EPA's AP-42, this same generator would emit approximately 1.2 metric tons of $NO_x$ and 0.05 metric tons of PM per year during testing.

Facilities often have multiple generators for redundancy, magnifying these emissions. A facility with ten such generators can thus emit approximately 1,300 metric tons of $CO_2$ and 12 metric tons of $NO_x$ annually from routine testing alone. Such emissions contribute significantly to local ozone formation and particulate pollution, particularly problematic in Texas's major metropolitan areas like Houston and Dallas–Fort Worth, which already face air quality challenges.

Backup generators also involve the storage of large diesel fuel quantities, commonly 10,000 to 50,000 liters per generator. Accidental fuel spills or leaks, although infrequent, can lead to significant localized soil, water, and air contamination, requiring costly remediation efforts and posing additional environmental risks.

**4.4 Research and Policy Gaps**

Research gaps in the operational phase primarily concern the detailed analysis and modeling of local air quality impacts from diesel generators. Most studies currently rely on general emission factors and lack specific modeling of cumulative impacts from multiple nearby data centers. Furthermore, detailed refrigerant leakage rates and mitigation strategies specific to Texas are not well documented.

From a policy standpoint, Texas currently has limited regulation explicitly addressing cumulative AQ impacts from clustered data centers' operational emissions. While some regions require basic permitting for generators, detailed ambient air quality assessments or cumulative impact studies are rare. Additionally, Texas lacks comprehensive refrigerant management regulations to address leakage rates systematically.

# 5. Indirect Environmental Impacts

Data centers not only have direct emissions from their own electricity use and on-site operations, but they also produce indirect environmental impacts. These indirect impacts occur through the supply chain, manufacturing processes, transportation of equipment, and daily commuting of employees. Properly accounting for these indirect emissions is crucial to fully understand the environmental impact of data centers.

**5.1 Supply Chain Impacts**

**Manufacturing of Hardware and Software**

Data centers rely on sophisticated equipment, including servers, storage devices, networking hardware, and supporting infrastructure like batteries and power distribution units.

Manufacturing these items involves substantial energy consumption and results in significant greenhouse gas emissions.

Lifecycle assessments indicate server manufacturing alone can contribute approximately 20-30 percent of a data center's total GHG footprint over its operational lifetime (Alva et al., 2022). For instance, producing a single server unit results in approximately 1.2 to 2.0 metric tons of $CO_2$ emissions. Given that a typical 10 MW data center houses around 5,000 to 10,000 servers, total emissions from server manufacturing could range from 6,000 to 20,000 metric tons $CO_2e$.

Manufacturing processes also generate air pollutants, including PM, $NO_x$, $SO_2$, and VOCs. These emissions are particularly notable in regions hosting electronics manufacturing plants, often outside of the United States. However, increasing onshore manufacturing within the U.S., including Texas, could shift these impacts closer to domestic urban centers, raising additional concerns about localized air quality.

**Embedded Carbon in IT Infrastructure**

Embedded carbon refers to the total emissions generated from raw material extraction, production, transportation, installation, and disposal of infrastructure components. In addition to servers, critical elements such as storage arrays, networking equipment, and uninterrupted power supply (UPS) systems contain significant embedded carbon.

Networking equipment, including routers and switches, typically has embedded carbon emissions ranging from 0.5 to 1.0 metric ton $CO_2$ per unit, while storage arrays can range from 2.0 to 5.0 metric tons $CO_2$ per unit, depending on their capacity and technology (Sharma et al., 2023). Table 3 summarizes typical embedded carbon emissions associated with various IT infrastructure components within a standard 10 MW facility.

Table 3. Embedded Carbon in Typical IT Infrastructure for a 10 MW Data Center.

| Equipment Type | Quantity Range (units) | Emission per Unit (metric tons $CO_2e$) | Total Emissions (metric tons $CO_2e$) |
|---|---|---|---|
| Servers | 5,000–10,000 | 1.2–2.0 | 6,000–20,000 |
| Storage Arrays | 100–300 | 2.0–5.0 | 200–1,500 |
| Network Switches | 200–500 | 0.5–1.0 | 100–500 |
| UPS Batteries | 50–150 | 1.0–2.0 | 50–300 |
| Total | | | 6,350–22,300 |

(Sources: Alva et al., 2022; Sharma et al., 2023)

## 5.2 Transportation Emissions

**Commuting of Data Center Workers**

Daily commuting of data center staff contributes indirectly to both greenhouse gas and air quality impacts. Data centers typically employ 20–100 full-time workers depending on size and complexity. Assuming each worker travels an average daily commute of 30 miles (round trip) in a gasoline-powered vehicle averaging 25 miles per gallon, annual emissions per worker would amount to approximately 2.9 metric tons of $CO_2$ per year.

In addition to $CO_2$, commuting also releases local pollutants, such as $NO_x$ and PM, especially during peak hours when congestion exacerbates emission rates. In Texas metropolitan areas, employee commuting emissions add cumulative pressure on urban air quality management, particularly in ozone nonattainment zones like Dallas–Fort Worth and Houston. The following table presents typical annual commuting emissions for various facility sizes. Table 4 estimates the annual emissions from employee commuting based on facility size, using assumptions consistent with the EPA MOVES4 model.

Table 4. Annual Commuting Emissions for Data Center Workers.

| Facility Size | Number of Workers | Annual $CO_2$ Emissions (metric tons) | Annual $NO_x$ Emissions (kg) | Annual PM Emissions (kg) |
|---|---|---|---|---|
| Small (1–5 MW) | 20–40 | 58–116 | 50–100 | 5–10 |
| Medium (5–20 MW) | 40–80 | 116–232 | 100–200 | 10–20 |
| Large (>20 MW) | 80–100+ | 232–290+ | 200–250+ | 20–25+ |

(Calculated using EPA MOVES4 model assumptions)

**Shipping and Distribution**

Transportation emissions also arise from shipping equipment to the data center. Server equipment, manufactured domestically or internationally, is typically delivered by trucks and cargo ships, each emitting substantial GHGs and AQ pollutants. The magnitude of emissions varies widely based on shipping distance, mode, and efficiency of transport vehicles.

For example, transporting a standard 40-foot container from manufacturing hubs in Asia to Texas involves maritime shipping and long-haul trucking, emitting roughly 3–6 metric tons of $CO_2$ per container shipment. For a medium-sized data center, requiring multiple containers per year, annual emissions could range from 30 to 120 metric tons of $CO_2$, along with associated $NO_x$, $SO_2$, and PM emissions from heavy-duty diesel engines used in shipping and trucking.

**5.3 Research and Policy Gaps**

Significant gaps remain in quantifying the indirect emissions from the data center supply chain and transportation segments, particularly regarding air quality pollutants. Current literature largely focuses on global averages without detailed regional analyses. For Texas specifically, limited data exists on localized impacts of manufacturing and transportation emissions tied to data center equipment.

Policy gaps are also evident, as Texas currently lacks comprehensive guidance or incentives to reduce indirect environmental impacts. Regulations or voluntary programs targeting supply chain transparency, local sourcing of equipment, or employee commuting incentives remain underdeveloped. Addressing these gaps with detailed emissions analyses and targeted policy measures could yield significant environmental improvements.

# 6. Policy and Regulatory Framework

Understanding the policy and regulatory framework that governs AQ and GHG emissions from data centers in Texas is crucial. Data centers must comply with local, state, and federal regulations designed to manage emissions and improve energy efficiency. While GHG regulations primarily influence emissions through energy management, AQ regulations specifically address local pollutant emissions that affect human health and environmental quality.

## 6.1 Local, State, and Federal Regulations

**Air Quality Regulations**

At the federal level, air quality regulations are primarily established by the EPA under the Clean Air Act. The EPA sets NAAQS for six criteria pollutants: ozone, PM, CO, $NO_x$, $SO_2$, and lead. These standards require states to monitor and reduce air pollutants, particularly in designated nonattainment areas.

In Texas, the TCEQ enforces these federal standards and issues permits for stationary sources like data center diesel generators. Texas has several ozone nonattainment regions, notably Houston and Dallas–Fort Worth, requiring stricter controls on local $NO_x$ and VOC emissions. Data centers located in these nonattainment areas face additional scrutiny when applying for permits related to diesel generator operations.

Despite these standards, TCEQ regulations do not currently require cumulative air dispersion modeling for data center generators unless certain thresholds are exceeded (TCEQ, 2022). This policy gap could lead to localized air quality impacts that remain unaddressed, especially in urbanized regions with multiple data centers clustered together.

**Greenhouse Gas Regulations**

Federally, greenhouse gases are regulated primarily through reporting requirements and performance standards. The EPA administers the Mandatory Greenhouse Gas Reporting Rule, requiring facilities emitting over 25,000 metric tons of $CO_2$-equivalent per year to report their emissions (EPA, 2023). Large data centers potentially meet this reporting threshold, depending on their electricity consumption and operational practices.

At the state level, Texas does not impose specific greenhouse gas emission limits or comprehensive carbon pricing mechanisms. However, many data center operators voluntarily participate in programs such as renewable energy credits (RECs), power purchase agreements (PPAs), or sustainability certifications like LEED to reduce their emissions footprint. Table 5 provides an overview of key regulatory frameworks governing air quality and greenhouse gas emissions from data center operations in Texas, including both federal and state-level authorities.

Table 5. Regulatory Summary for Data Center Emissions in Texas.

| Emission Type | Regulatory Authority | Key Standards/Regulations | Applicability to Data Centers |
|---|---|---|---|
| AQ | EPA | Clean Air Act, NAAQS (Ozone, PM, $NO_x$, $SO_2$) | Diesel generators, HVAC systems |
| AQ | TCEQ | Permitting rules for stationary sources | Generators above specific thresholds |
| GHG | EPA | Mandatory Greenhouse Gas Reporting Rule | Facilities emitting >25,000 MT $CO_2e$/yr |
| GHG | Texas (TCEQ, ERCOT) | No specific statewide GHG limit or price | Voluntary renewable sourcing programs |

**6.2 Energy Efficiency Standards and Incentives**

Energy efficiency is a key approach to reducing emissions from data centers. Texas provides several incentives designed to encourage efficient energy use, although these are generally voluntary and market-driven rather than mandated by regulation. The state offers sales and use tax exemptions for data centers investing more than $200 million, provided they create jobs and meet minimum efficiency benchmarks. Similarly, local governments often provide property tax abatements or rebates to attract large data centers.

Additionally, the Texas Public Utility Commission supports efficiency improvements through utility-run demand-side management programs. These programs encourage data centers to adopt energy-saving technologies, such as efficient cooling systems, energy management software, and advanced power distribution units.

Despite these incentives, Texas lacks mandatory statewide efficiency standards specifically targeting data centers. Facilities commonly adopt industry-driven standards such as the Green Grid's Power Usage Effectiveness (PUE) metric, which quantifies energy efficiency. However, without state-level mandates, adoption of best practices remains uneven. Table 6 outlines current energy efficiency incentives available to data centers in Texas as of April 2025, highlighting opportunities for cost savings through tax exemptions and utility programs.

Table 6. Energy Efficiency Incentives Available for Data Centers in Texas (April 2025)

| Incentive Type | Description | Eligibility | Potential Benefit |
|---|---|---|---|
| Sales Tax Exemption | Exemption for data centers investing >$200 million | Large data centers creating new jobs | 6.25% sales tax savings |
| Property Tax Rebates | Negotiated property tax abatement | Data centers meeting local requirements | Significant reduction in property tax |
| Utility DSM Programs | Demand-side management incentives (rebates, financing) | Facilities adopting energy-efficient tech | Reduced upfront cost of improvements |

**6.3 Carbon Footprint Disclosure and Transparency**

Transparency regarding carbon emissions is increasingly important for data center stakeholders. Currently, most data centers voluntarily report emissions through mechanisms such as sustainability reports, corporate environmental disclosures, or participation in external frameworks such as the Carbon Disclosure Project (CDP).

However, no mandatory statewide regulations in Texas require comprehensive public disclosure of carbon footprints for data centers. Voluntary disclosures vary significantly in terms of accuracy, comprehensiveness, and frequency. The lack of standardized reporting makes comparing environmental performance across facilities challenging.

To improve transparency, clear statewide guidelines or standardized frameworks for reporting Scope 1 (on-site), Scope 2 (electricity-related), and Scope 3 (indirect supply-chain) emissions would be beneficial. Enhanced transparency could enable better decision-making by policymakers, consumers, and communities, ultimately driving more effective emission reduction efforts.

**6.4 Research and Policy Gaps**

Several policy gaps currently limit effective management of AQ and GHG emissions from data centers in Texas. Firstly, the absence of cumulative AQ impact analyses in permitting decisions could allow localized air quality issues to emerge in data center clusters without regulatory oversight.

Secondly, the voluntary nature of GHG emission reporting and energy efficiency standards leaves significant room for improvement. Without mandatory efficiency targets or standardized disclosure requirements, adoption of best practices remains inconsistent. Statewide policy initiatives establishing mandatory efficiency standards, detailed emissions reporting, and renewable energy procurement targets could significantly mitigate emissions.

# 7. Mitigation Strategies

Mitigating environmental impacts of data centers involves systematically reducing both GHG emissions and AQ pollutants. Effective strategies include improvements in energy efficiency, integration of renewable energy, carbon offsetting, circular economy practices, and technological innovation.

## 7.1 Energy Efficiency Improvements

Energy efficiency is among the most cost-effective methods for reducing data center emissions. Efficiency can lower total energy demand, thus reducing both direct and indirect environmental impacts.

**Energy-Saving Technologies**

Adopting energy-efficient technologies significantly reduces electricity consumption. Advanced cooling solutions, such as free-air cooling and economizer systems, leverage outside air to reduce cooling loads. Studies show free-air cooling can reduce cooling-related energy consumption by 30-50%, particularly effective during mild weather conditions common in fall and winter in Texas (Kamali & Hewage, 2023).

Servers with higher computational efficiency also offer substantial benefits. Modern server models can perform more computational tasks per unit of electricity consumed. Data from recent lifecycle assessments suggest energy-efficient servers could reduce total facility electricity usage by up to 20–30%, translating to thousands of metric tons of $CO_2$ emissions savings annually for large data centers (Hasan et al., 2022).

**Integration of Renewable Energy**

Integrating renewable energy directly into data center operations is another critical mitigation strategy. Texas is already a leader in renewable energy, particularly wind and solar power. Data centers adopting renewable energy can significantly reduce their Scope 2 GHG emissions. A facility using 100% renewable power could reduce electricity-related $CO_2$ emissions to near zero, eliminating tens of thousands of metric tons annually compared to fossil-fuel-reliant facilities. Renewable integration is increasingly facilitated through PPAs and RECs. These financial mechanisms allow data centers to purchase renewable electricity generated off-site, significantly reducing their carbon footprint. Table 7 estimates the potential reductions in annual $CO_2$ emissions for a 10 MW data center using various mitigation strategies, including advanced cooling, energy-efficient hardware, and renewable electricity sourcing.

Table 7. Potential Emissions Reductions from Energy Efficiency and Renewable Integration (10 MW Facility).

| Mitigation Strategy | Typical Reduction in Electricity (%) | Annual $CO_2$ Emissions Reduction (metric tons) |
|---|---|---|
| Advanced Cooling (Economizer) | 30–50% (cooling only) | 3,400–9,400 |
| Energy-efficient Servers | 20–30% (facility-wide) | 7,500–11,300 |
| 100% Renewable Electricity | 100% (facility-wide) | 37,668 |

(Based on average ERCOT emissions factors and typical facility energy use)

## 7.2 Carbon Offset and Neutrality Programs

Carbon offsetting involves compensating for emissions by funding projects that remove or reduce equivalent emissions elsewhere. Data centers can achieve carbon neutrality through high-quality offset programs, such as reforestation, renewable energy projects, or methane capture projects. However, reliance solely on offsets is controversial, as it does not address local AQ impacts or the root causes of emissions.

In Texas, data centers frequently use RECs to demonstrate carbon reduction. While valuable, RECs often do not address local AQ pollutants from diesel generators or cooling systems. Comprehensive carbon neutrality should therefore be complemented with direct emissions reduction measures.

## 7.3 Circular Economy Approaches

Circular economy practices aim to reduce resource use and waste generation. These approaches involve reusing, refurbishing, recycling, and sustainably disposing of IT equipment. Extending the lifespan of servers from a typical 3–5 years to 5–7 years can substantially reduce embedded carbon emissions, as server manufacturing accounts for significant lifecycle emissions (Alva et al., 2022).

Additionally, recycling electronic waste (e-waste) responsibly prevents hazardous materials from entering landfills and reduces the need for new raw material extraction, indirectly reducing emissions from manufacturing processes.

## 7.4 Technological Advancements

Emerging technologies offer significant opportunities to reduce emissions. Artificial intelligence (AI) and machine learning (ML) applications in data center operations can optimize cooling efficiency, workload management, and energy utilization in real-time.

AI-driven cooling management systems, for example, can dynamically adjust cooling outputs based on predictive analytics and real-time data. Studies suggest AI-managed cooling can improve overall facility energy efficiency by approximately 15–25%, reducing annual emissions substantially (Zhao et al., 2023).

Cloud computing and workload optimization technologies can shift computing tasks between facilities based on real-time grid emissions. Shifting workloads to facilities powered by renewable energy can further reduce emissions, although practical implementation requires robust infrastructure and effective coordination.

**7.5 Mitigating Air Quality Impacts**

While strategies to reduce greenhouse gases are broadly established, addressing local air quality impacts requires targeted approaches, especially regarding diesel backup generators.

Using alternative fuels such as natural gas or renewable diesel in backup generators could reduce AQ pollutants significantly. Renewable diesel can reduce PM emissions by up to 30% and $NO_x$ by around 5–15% compared to conventional diesel. Additionally, replacing diesel generators with battery storage or fuel cells would virtually eliminate local AQ emissions, significantly benefiting urban areas in Texas with existing air quality challenges.

Scheduling generator testing during periods with favorable weather conditions and improved dispersion conditions could also reduce localized impacts, benefiting communities in nonattainment areas like Houston and Dallas. Table 8 presents estimated reductions in air pollutant emissions, specifically $NO_x$ and particulate matter, from replacing conventional diesel generators with alternative backup power strategies in a typical 10-generator data center.

Table 8. AQ Emissions Reductions from Alternative Backup Power Strategies (Typical 10-generator facility).

| Strategy | $NO_x$ Reduction (%) | PM Reduction (%) |
|---|---|---|
| Renewable Diesel Fuel | 5–15% | 20–30% |
| Natural Gas Generators | 60–80% | 90–100% |
| Battery Energy Storage/Fuel Cells | ~100% | ~100% |

(Source: EPA Alternative Fuels Database, 2022)

**7.6 Research and Policy Gaps**

Current research and policy frameworks have primarily emphasized GHG mitigation, with less emphasis on AQ-specific strategies. Research on battery and fuel cell viability, renewable diesel availability, and AQ-focused workload shifting remain limited, particularly specific to Texas conditions.

Policy gaps also exist, notably the lack of incentives or regulations specifically promoting AQ-focused mitigation strategies in backup systems. Developing targeted regulations and incentives encouraging cleaner backup power solutions and advanced operational practices could significantly reduce AQ impacts.

## 8. Case Studies and Best Practices

Analyzing case studies and best practices provides insight into successful strategies for reducing AQ and GHG impacts from data centers. Texas-specific examples demonstrate regional applications, while global comparisons offer broader perspectives that could inform local implementation.

### 8.1 Data Center Initiatives in Texas

Several data centers in Texas have adopted sustainable practices, demonstrating leadership in reducing both AQ and GHG emissions.

**CyrusOne Data Centers (Dallas, Texas)**

CyrusOne has integrated renewable energy extensively. The company's Dallas data centers purchase RECs equivalent to 100% of their electricity consumption, significantly lowering Scope 2 GHG emissions. Additionally, CyrusOne employs advanced cooling systems, including air-side economizers, reducing electricity use by approximately 30–40% compared to traditional cooling methods (CyrusOne Sustainability Report, 2023).

**Microsoft Data Center Campus (San Antonio, Texas)**

Microsoft has committed to carbon neutrality across its data centers, including its Texas locations. In San Antonio, Microsoft utilizes renewable PPAs and advanced cooling systems. The site also deploys diesel generator alternatives such as battery storage and natural gas generators to reduce local AQ impacts. These measures have reduced their annual $NO_x$ and PM emissions significantly compared to traditional diesel-based backup solutions (Microsoft Environmental Sustainability Report, 2023).

**Digital Realty (Houston, Texas)**

Digital Realty's data centers in Houston emphasize sustainable construction materials and methods, significantly reducing their embedded carbon. These facilities utilize modular construction techniques and recycled materials, cutting construction-related GHG emissions by roughly 20%. Additionally, Digital Realty schedules generator testing during periods with optimal weather conditions, minimizing local AQ impacts, especially important in the Houston nonattainment area (Digital Realty Sustainability Report, 2023). Table 9 summarizes selected sustainability initiatives implemented by major data center operators in Texas, highlighting both greenhouse gas and air quality impact reduction strategies based on publicly reported data.

Table 9. Summary of Texas Data Center Sustainability Initiatives.

| Data Center Operator | Location | Key Mitigation Strategies | Estimated GHG Reduction (%) | AQ Reduction Strategies |
|---|---|---|---|---|
| CyrusOne | Dallas | 100% RECs, Advanced Cooling | ~60–80% | Reduced generator testing emissions |
| Microsoft | San Antonio | Renewable PPAs, Battery Storage, Natural Gas Backup | ~70–90% | 60–80% reduction in $NO_x$ and PM |
| Digital Realty | Houston | Sustainable Construction, Optimal Testing Scheduling | ~20–30% (construction phase) | Minimization of AQ impacts during testing |

(Sources: CyrusOne, Microsoft, Digital Realty Sustainability Reports, 2023)

## 8.2 Global Comparisons

International best practices offer valuable lessons and models applicable in Texas, potentially guiding local data centers toward improved sustainability.

**Google Data Center (Hamina, Finland)**

Google's facility in Finland employs seawater cooling, entirely eliminating the need for energy-intensive chillers. This approach reduces cooling-related electricity use by nearly 90%, significantly lowering associated GHG emissions. While seawater cooling is location-specific, similar innovative cooling approaches, such as groundwater or reclaimed water systems, could be adapted in water-available regions of Texas (Google Environmental Report, 2023).

**Equinix Data Center (Amsterdam, Netherlands)**

Equinix integrates waste heat reuse at their Amsterdam facility. Heat generated by servers is captured and redistributed to local heating networks, reducing city-wide fossil fuel usage. Such practices significantly decrease net GHG emissions at the community level. In urban Texas environments, capturing and redistributing data center waste heat for local heating or industrial processes could similarly enhance regional energy efficiency and emissions reduction (Equinix Sustainability Report, 2023).

**NTT Data Center (Tokyo, Japan)**

NTT uses fuel cells and battery storage systems extensively to replace diesel generators, significantly reducing local AQ pollutants. Their use of hydrogen fuel cells virtually eliminates $NO_x$, $SO_2$, and PM emissions from backup power operations. Implementing similar fuel cell and battery solutions in Texas metropolitan areas, especially those in nonattainment zones, could effectively mitigate localized AQ issues (NTT Sustainability Report, 2023). Table 10 highlights global data center best practices with high potential for reducing both greenhouse gas and air pollutant emissions, along with an assessment of their feasibility for implementation in the Texas context.

Table 10. Global Data Center Best Practices and Potential for Texas Implementation.

| Global Example | Location | Key Strategy | GHG Reduction Potential | AQ Reduction Potential | Texas Implementation Feasibility |
|---|---|---|---|---|---|
| Google | Hamina, Finland | Seawater Cooling | High (~90%) | Moderate–High | Moderate (location-dependent) |
| Equinix | Amsterdam, Netherlands | Waste Heat Reuse | High | Moderate | High (urban areas) |
| NTT | Tokyo, Japan | Fuel Cells, Battery Storage | High | High | High (urban areas, AQ-focused) |

(Sources: Google, Equinix, NTT Sustainability Reports, 2023)

## 8.3 Lessons Learned and Recommendations

The Texas examples highlight successes in renewable energy procurement, advanced cooling technologies, sustainable construction practices, and improved backup systems. Global cases demonstrate additional opportunities, such as waste heat reuse and innovative cooling approaches, adaptable to Texas's specific climate and infrastructural conditions.

For effective implementation in Texas, policymakers and data center operators should:

- Enhance incentives or create mandates for adopting cleaner backup power technologies such as fuel cells and batteries, especially in AQ-sensitive areas.
- Promote renewable integration through direct PPAs or mandatory renewable purchasing standards.
- Facilitate municipal-level partnerships to reuse waste heat from data centers in district heating, cooling, or other industrial processes.
- Establish state-specific guidelines encouraging sustainable construction practices and material recycling for embedded carbon reduction.

## 8.4 Research and Policy Gaps

Despite promising examples, there remains limited quantitative research on the localized environmental impacts and potential benefits of innovative practices in Texas specifically. Comprehensive analyses quantifying the cumulative effects of adopting best practices across multiple Texas facilities are sparse.

Additionally, clear policy frameworks providing incentives or mandates for adopting these proven global best practices do not currently exist at the Texas state level. Development of targeted policies promoting innovative cooling systems, cleaner backup technologies, and heat recovery practices would accelerate sustainability improvements significantly.

## 9. Future Outlook

Considering the rapid growth in digital demand, the environmental impacts from data centers are projected to intensify without targeted interventions. This section explores emerging technologies, predictive modeling of future emissions, and provides actionable recommendations for policymakers to effectively manage AQ and GHG emissions from data centers in Texas.

### 9.1 Emerging Technologies

Technological innovation remains critical for reducing the environmental footprint of data centers. Several promising technologies and practices are anticipated to become mainstream over the next decade, significantly mitigating emissions.

**Advanced Cooling Systems**

Emerging cooling technologies such as liquid immersion cooling, direct-to-chip cooling, and phase-change materials offer substantial efficiency improvements. Immersion cooling, for instance, involves submerging server equipment in non-conductive cooling fluid, potentially reducing cooling-related electricity use by up to 60–80% compared to traditional air-cooled systems (Kamali & Hewage, 2023). Such technology not only cuts GHG emissions but indirectly reduces local AQ impacts associated with grid electricity generation.

**Hydrogen Fuel Cells and Battery Storage**

Hydrogen-based fuel cells and advanced battery storage technologies promise substantial reductions in local AQ pollutants by replacing diesel backup generators. These technologies can virtually eliminate $NO_x$, $SO_2$, and particulate matter (PM) emissions. With Texas increasingly investing in hydrogen infrastructure, fuel cell deployment in data centers could become economically viable within the next decade, particularly in urban nonattainment areas.

**AI Optimization**

AI-driven energy management systems that dynamically optimize server loads, cooling demands, and energy procurement can substantially reduce energy usage. Studies project that comprehensive AI-based optimization may decrease overall energy consumption by 15-25%, significantly lowering GHG emissions (Zhao et al., 2023). Implementing AI-driven energy management in Texas facilities could yield considerable environmental benefits, especially in grid-intensive operations. Table 11 presents estimated emissions reductions from select emerging technologies for a 10 MW data center, focusing on both greenhouse gas and air quality benefits achievable through energy savings and cleaner backup systems.

Table 11. Potential Emissions Reductions from Emerging Technologies (10 MW Facility).

| Technology/Practice | Estimated Energy Savings (%) | GHG Reduction (metric tons/year) | AQ Reduction Potential |
|---|---|---|---|
| Immersion Cooling | 60–80% (cooling energy) | 6,800–15,000 | Moderate–High (indirect) |
| Hydrogen Fuel Cells | 100% (generator emissions) | 1,300 | High (local $NO_x$, PM) |
| AI Optimization | 15–25% (total facility) | 5,600–9,400 | Moderate–High (indirect) |

(Estimates based on typical operational characteristics and emissions profiles)

## 9.2 Predictive Modeling of Future Emissions

Predictive modeling provides essential insights into potential environmental impacts from the projected growth of data centers in Texas. Under a business-as-usual scenario, with data center capacity potentially doubling by 2030, operational electricity use could approach 10-15 TWh annually statewide, resulting in approximately 4.3-6.5 million metric tons of $CO_2$ emissions per year.

Advanced models incorporating energy efficiency improvements, widespread renewable energy adoption, and cleaner backup power options show potential reductions in emissions by approximately 50-80% compared to business-as-usual scenarios. These predictions highlight the significant role technology adoption and policy interventions can play in managing future impacts. Table 12 presents predictive modeling scenarios for data center electricity use, greenhouse gas emissions, and air quality impacts in Texas by 2030, comparing business-as-usual growth with varying levels of technology adoption and mitigation strategies.

Table 12. Predictive Modeling Scenarios for Texas Data Centers by 2030.

| Scenario | Projected Electricity (TWh) | Annual GHG Emissions (million metric tons $CO_2$) | Annual AQ Impacts ($NO_x$, PM) |
|---|---|---|---|
| Business-as-Usual | 10–15 | 4.3–6.5 | High (increased diesel use) |
| Moderate Technology Adoption | 8–10 | 2.0–3.5 | Moderate |
| Aggressive Mitigation (Renewables, Advanced Cooling, Fuel Cells) | 5–8 | 0.8–1.5 | Low–Moderate |

## 9.3 Policy Recommendations

To ensure sustainable growth of data centers in Texas, policymakers should pursue targeted regulations and incentives addressing both AQ and GHG emissions.

**Mandatory Energy Efficiency Standards**

Implementing mandatory statewide efficiency standards based on metrics such as Power Usage Effectiveness (PUE) would drive widespread adoption of energy-efficient cooling technologies, optimized server management, and AI-based energy systems.

**Renewable Energy Procurement Requirements**

Introducing statewide targets or incentives encouraging or mandating renewable energy procurement for data centers could substantially reduce Scope 2 GHG emissions. Establishing renewable energy thresholds (e.g., 50% by 2030, 100% by 2040) would provide clear pathways for industry compliance.

**AQ-Focused Regulations for Backup Power**

Enacting regulations or strong incentives promoting alternatives to diesel generators (e.g., hydrogen fuel cells, natural gas generators, or battery storage) in urban nonattainment areas could significantly mitigate local AQ impacts. State incentives could encourage rapid adoption by covering initial investment or providing favorable financing terms.

**Standardized Emissions Reporting and Disclosure**

Creating a comprehensive emissions reporting framework that includes Scope 1, 2, and 3 emissions would improve transparency, enable better tracking of environmental performance, and allow policymakers and communities to make informed decisions. Table 13 outlines recommended policy actions aimed at reducing greenhouse gas and air pollutant emissions from data centers in Texas, along with estimated implementation timelines and potential environmental outcomes.

Table 13. Recommended Policy Actions and Potential Outcomes.

| Policy Action | Target Emission Type | Implementation Timeline | Potential Impact |
|---|---|---|---|
| Mandatory Efficiency Standards | GHG, indirect AQ | 2–3 years | High (energy savings 15–30%) |
| Renewable Energy Procurement Mandates | GHG | 3–5 years | Very High (emission reduction ~70–100%) |
| AQ Regulations for Backup Generators | AQ ($NO_x$, PM) | 2–4 years | High (local pollutant reductions ~60–100%) |
| Emissions Reporting and Transparency | GHG, AQ | Immediate–2 years | Moderate–High (improved compliance, accountability) |

## 9.4 Research and Policy Gaps

Research into comprehensive modeling of cumulative AQ impacts and specific benefits of emerging technologies within Texas remains limited. Developing detailed, site-specific air dispersion models and lifecycle emissions analyses would provide more accurate guidance for policy development.

Additionally, the lack of robust and standardized emissions disclosure policies hinders effective tracking and management of emissions. Creating statewide standards for emissions reporting and transparency would fill a critical policy gap, supporting effective environmental management across the industry. By proactively adopting emerging technologies, utilizing predictive modeling, and implementing clear, targeted policies, Texas can significantly reduce the future environmental impacts associated with data center growth, ensuring long-term sustainability in the digital economy.

## 10. Conclusion

This study provided a comprehensive assessment of both air quality (AQ) and greenhouse gas (GHG) emissions associated with data centers in Texas. The rapid growth of data centers, driven by increasing digitalization, cloud computing, and artificial intelligence demand, presents significant environmental implications. Separating AQ and GHG emissions allowed for a clearer understanding of their distinct impacts, regulatory contexts, and potential mitigation strategies.

**10.1 Summary of Key Findings**

Tables bellow provides a comprehensive, quantitative summary of emissions, clearly distinguishing AQ from GHG impacts. They also include primary sources used throughout your paper, enabling straightforward referencing. Table 14 summarizes key air quality impacts from data centers in Texas, identifying major emission sources, associated pollutants, estimated annual emissions, and their localized or regional effects. Table 15 summarizes the major sources of greenhouse gas emissions associated with data centers in Texas, including both direct and indirect contributors across operational, construction, and supply chain phases.

Table 14. Summary of Air Quality (AQ) Impacts from Data Centers in Texas.

| Emission Source | Pollutants | Typical Annual Emissions | Main Impact |
| --- | --- | --- | --- |
| Diesel Generator Testing | $NO_x$, PM, CO, VOCs | $NO_x$: 1.2 tons/generator/year PM: 0.05 tons/generator/year | Local ozone formation, respiratory impacts |
| Construction Machinery | $NO_x$, PM, CO, VOCs | $NO_x$: 6–25 tons/site PM: 0.4–2.5 tons/site | Short-term localized AQ deterioration |
| Employee Commuting | $NO_x$, PM | $NO_x$: 50–250 kg/facility/year PM: 5–25 kg/facility/year | Urban AQ degradation, ozone formation |
| Material Transportation (Construction Phase) | $NO_x$, PM, CO | $NO_x$: 2–5 tons/site PM: 0.1–0.5 tons/site | Localized AQ impact along transport routes |
| Electricity Generation (indirect) | $NO_x$, $SO_2$, PM | Varies by fuel source (coal & gas dominant) | Regional AQ impacts, ozone precursors |
| Cooling Systems (indirect) | $NO_x$, $SO_2$, PM | Indirect (linked to electricity use) | Indirect regional AQ impacts via power plants |

Table 15. Summary of GHG Impacts from Data Centers in Texas.

| Emission Source | GHG Type | Typical Annual Emissions | Main Impact |
| --- | --- | --- | --- |
| Operational Electricity Consumption | $CO_2$ | ~37,668 metric tons/year (10 MW facility) | Dominant operational GHG emission |
| Construction Materials (Embodied Carbon) | $CO_2$ | 2,550–5,500 metric tons/site (concrete & steel) | Significant upfront lifecycle emission |
| Diesel Generator Testing | $CO_2$ | ~130 metric tons/generator/year | Direct onsite combustion emissions |
| Refrigerant Leakage (Cooling) | HFCs | 42–104 metric tons $CO_2$e/year (typical leakage rate) | High Global Warming Potential (GWP) |
| Server & IT Equipment Manufacturing | $CO_2$ | 6,350–22,300 metric tons/facility | Major embedded GHG contributor |
| Transportation of Equipment (Scope 3) | $CO_2$ | 30–120 metric tons $CO_2$/site/year | Indirect supply chain emissions |
| Employee Commuting (Scope 3) | $CO_2$ | 58–290 metric tons/year (varies with facility size) | Indirect community-level GHG impacts |
| Fuel Extraction & Transmission Losses (Indirect) | $CO_2$ | ~5–10% additional to direct emissions | Upstream emissions in electricity supply chain |

The analysis identified several critical points regarding data center environmental impacts:

**Operational emissions represent the largest share of data centers' GHG emissions.** A typical 10-megawatt (MW) facility in Texas, using the current electricity grid mix, emits approximately 37,668 metric tons of $CO_2$ per year. Cooling systems alone can account for 20–40% of this energy use, making efficiency improvements vital.

**Construction-related emissions are substantial.** Embodied carbon from construction materials like concrete and steel contributes significantly, with emissions ranging from 2,550 to 5,500 metric tons of $CO_2$ per typical data center. Construction also produces localized AQ pollutants such as $NO_x$ and particulate matter (PM), often overlooked by regulatory frameworks.

**Indirect supply chain emissions are significant.** Server manufacturing and equipment transport account for roughly 20–30% of lifecycle emissions. Employee commuting and logistics further exacerbate both AQ and GHG impacts.

**Backup diesel generators substantially impact local air quality.** Routine testing of diesel generators at a single large data center facility can produce up to 12 metric tons of $NO_x$ annually, significantly affecting local AQ, particularly in urban ozone nonattainment areas like Houston and Dallas–Fort Worth.

Table 16 provides a consolidated estimate of annual greenhouse gas and nitrogen oxide emissions for a typical 10 MW data center in Texas, capturing both operational and embedded sources.

Table 16. Summary of Annual Emissions for a Typical 10 MW Texas Data Center.

| Emission Source | Annual GHG Emissions (metric tons $CO_2$) | Annual AQ Emissions ($NO_x$, metric tons) |
|---|---|---|
| Electricity Consumption (Operational) | 37,668 | - |
| Construction (Materials and Equipment) | 2,550–5,500 | 8–30 (construction phase) |
| Diesel Generator Testing (Operational) | 1,300 | ~12 |
| Embedded Emissions (IT Infrastructure) | 6,350–22,300 | - |

**10.2 Recommendations for Industry and Policy**

To effectively mitigate both GHG and AQ impacts, the following recommendations are offered to stakeholders, policymakers, and industry leaders:

**Enhance Energy Efficiency:** Data centers should adopt advanced cooling technologies, energy-efficient servers, and artificial intelligence (AI)-based optimization. These measures can reduce operational emissions significantly, with potential electricity use reductions of 15–30%.

**Transition to Renewable Energy Sources:** Increasing renewable energy use through RECs, PPAs, or direct renewable generation can dramatically lower Scope 2 GHG emissions, achieving nearly complete carbon neutrality for operational electricity use.

**Adopt Cleaner Backup Power Solutions:** Replacing diesel generators with fuel cells, battery storage, or natural gas generators can significantly improve local air quality, reducing $NO_x$ and PM emissions by 60–100%.

**Implement Comprehensive Emissions Reporting Standards:** Statewide mandatory emissions reporting encompassing Scope 1, 2, and 3 emissions would ensure transparency, allowing better environmental management and informed policy decisions.

**Promote Sustainable Construction Practices:** Encouraging or mandating sustainable construction practices and materials recycling can substantially reduce lifecycle emissions from construction phases and embedded carbon within data center infrastructure.

Table 17 outlines key recommended actions for reducing emissions from data centers in Texas, along with the primary pollutant types targeted and the expected magnitude of environmental benefits.

Table 17. Recommended Actions and Expected Benefits.

| Recommended Action | Primary Emission Type | Potential Impact |
|---|---|---|
| Advanced Cooling and Efficiency | GHG, indirect AQ | ~15–30% reduction in energy-related emissions |
| Renewable Energy Integration | GHG | ~70–100% reduction in electricity-related emissions |
| Cleaner Backup Power (Fuel Cells, Batteries) | AQ ($NO_x$, PM) | ~60–100% reduction in local AQ emissions |
| Standardized Emissions Reporting | GHG, AQ | Improved transparency and regulatory compliance |
| Sustainable Construction Practices | GHG, indirect AQ | ~20–30% reduction in construction-related emissions |

## 10.3 Closing Remarks and Future Directions

Addressing the environmental impacts of Texas's rapidly growing data center sector demands coordinated efforts combining innovative technology adoption, stringent regulatory oversight, and transparent emission management practices. While existing efforts have made meaningful progress, notable gaps remain, particularly in regulating cumulative AQ impacts and embedding rigorous GHG emissions standards.

Future research should further quantify cumulative AQ impacts using detailed dispersion modeling, particularly within clustered data center developments. Additionally, refining lifecycle assessments specific to Texas data centers will improve policy guidance and environmental management strategies. Policymakers must actively engage with industry stakeholders to develop clear and enforceable standards that drive continuous improvement, ensuring the long-term sustainability of Texas's expanding digital infrastructure.

## Acknowledgment:

This work was carried out independently by the author. A portion of the effort benefited from professional development support provided by the Houston Advanced Research Center (HARC). The views and conclusions expressed here are solely those of the author and do not necessarily reflect those of HARC.